\documentclass[twocolumn,secnumarabic,amssymb, nobibnotes, aps, prd]{revtex4-2}

\usepackage{graphicx, bm}
\begin{document}

\title{Evaluation on Genetic Algorithms as an optimizer of Variational Quantum Eigensolver(VQE) method
}%

\author{Hikaru Wakaura}%
\email[Quantscape: ]{
hikaruwakaura@gmail.com}
\affiliation{QuantScape Inc. QuantScape Inc., 4-11-18, Manshon-Shimizudai, Meguro, Tokyo, 153-0064, Japan}

\author{Takao Tomono}

\affiliation{ Digital Innovation Div. TOPPAN Inc. , 1-5-1, Taito, Taito, Tokyo, 110-8560, Japan}
\email[TOPPAN: ]{takao.tomono@ieee.org}

\author{Shoya Yasuda}
\affiliation{Vignette $\&$ Clarity Inc., 5-21, Shirahataminami-cho, Kanagawa-ku, Yokohama, Kanagawa, 221-0073, Japan}
\email[Vigne \& Cla: ]{yasuda@vigne-cla.com}

\date{October 2021}%

\begin{abstract}

Variational-Quantum-Eigensolver(VQE) method on a quantum computer is a well-known hybrid algorithm to solve the eigenstates and eigenvalues that uses both quantum and classical computers.
This method has the potential to solve quantum chemical simulation including polymer and complex optimization problems that are never able to be solved in a realistic time.
Though they are many papers on VQE, there are many hurdles before practical application.
Therefore, we tried to evaluate VQE methods with Genetic Algorithms(GA).
In this paper, we propose the VQE method with GA.
We selected ground and excited-state energy on hydrogen molecules as the target because there are many local minimum values on excited states though the molecular structure is extremely simple.
Therefore it is not easy to find the energy of states.
We compared the GA method with other methods from the viewpoint of log error of the ground, triplet, singlet, and doubly excited state energy value.
As a result, we denoted that the BFGS method has the highest accuracy. We thought that rcGA used as an optimization for the VQE method was proved disappointing. The rcGA does not show an advantage compared to other methods. we suggest that the cause is due to initial convergence.
In the future, we want to try to introduce Genetic Algorithms then local search.

\end{abstract}

\maketitle
\tableofcontents

\maketitle

\section{Introduction}\label{1}
The Variational Quantum Eigensolver (VQE) \cite{PhysRevX.6.031007} method has been developed by Dr. Alan Aspuru-Guzik and his group has introduced the world in 2011 in their paper \cite{doi:10.1146/annurev-physchem-032210-103512}. In 2016, several cloud services using quantum computers with up to five qubits have emerged. In 2016, many cloud services using quantum computers with up to five qubits appeared, sparking the momentum to calculate molecular wavefunctions on quantum computers, and the hydrogen molecule \cite{PhysRevX.8.011021}, water molecule \cite{2019arXiv190210171N}, and LiH\cite{2017Natur.549..242K}. Their wave functions can now be calculated by quantum computers.

This movement accelerated the development of quantum hardware and server lack sized Ion-trap quantum computer by the group of the University of Innsbruck is released. The development of quantum computers using Nitro-Vacancy center qubits also accelerated. Besides, IBM declared that they will release a 1000-qubit quantum computer.
Today, not only superconducting qubits but also Ion-Trap quantum computers with high computational fidelity are available in the cloud.
As the number of available qubits increases, the number of electron orbitals that can be handled increases, hence it is expected that in the not too distant future, anyone will be able to solve the wavefunction of a large molecule with more than ten atoms. However, solving the wavefunction of large molecules is not an easy problem. A quantum computer generates noise, which causes decoherence of qubits. Therefore, the calculation must be completed within the coherence time. Furthermore, the more qubits required for a calculation, the lower the accuracy of the calculation. If quantum error correction methods are implemented in a quantum computer, the lifetime of the qubits will be longer. Even a two-dimensional surface topological code, which is assumed to be the most practical error correction method, requires $4n$ error-correcting qubits for $n$ data qubits\cite{PhysRevA.85.060301}. Current quantum computers are not capable of implementing quantum error correction methods. Hence, current quantum computers are called Noisy-Intermediate-Scale-Quantum (NISQ) computers. Currently, only quantum algorithms that can be used under these constraints are used, and the VQE method is one of them.
Several software packages have already been released for cloud computing, all of which can use VQE \cite{qiskit}\cite{pyquil}\cite{cirq}\cite{blueqat}\cite{renomq}.
The more bits that are needed for the calculation, the more variables are needed, which makes the calculation more time-consuming.

For the calculation of the ground state, the Unitary-Coupled-Cluster (UCC) Ansatz is well known for the calculation of the wave function in the VQE method.
This Ansatz requires $O(n^2n_v^2 / 4)$ variables for the number of occupied orbitals $n$ and the number of unoccupied orbitals $n_v$, even in the case of UCCSD\cite{2018PhRvA..98b2322B}, where excitations are limited to two electrons. In addition, the calculation of excited states requires the accuracy to find the correct allowed excited states from among those that are physically not allowed.

The optimization methods used in VQE include the Powell's method\cite{2018PhRvA..98b2322B}, the Nelder-Mead method\cite{McClean_2016}, the Broyden-Fletcher-Goldfarb-Shanno (BFGS)\cite{2018PhRvA..98b2322B} method and Conjugate Gradient (CG) method\cite{BOUWMEESTER2015276}. Powell's method is a search method using directional vectors. The Nelder-Mead method searches for the minimum value of a function while moving a vector in an n-dimensional direction consisting of n+1 vertices. The Nelder-Mead method searches for the minimum value of a function while moving a vector in an n-dimensional direction with n+1 vertices, using three different methods: reflection, expansion, and contraction.
The BFGS method is a kind of pseudo-Newtonian method that searches for the stopping point of a function whose gradient becomes zero by differentiating it twice. The CG method is also a kind of pseudo-Newton method, which searches in a straight line from the starting point to the extreme value of the goal.

Bayesian methods, on the other hand, are based on the concept of Bayesian probability and infer things to be inferred from observed facts in a probabilistic sense. Therefore, it is recognized as a method suitable for obtaining a globally optimal solution without falling into a locally optimal solution.

Genetic Algorithm (GA) is a meta-heuristic algorithm that mimics the process of biological evolution. In this algorithm, we prepare a large number of individuals whose parameters to be optimized are represented by genes. The algorithm selects multiple parents from a diverse population, mates them to produce offspring, and repeats the generation cycle. While GA is easy to implement, it requires special consideration for each problem, such as how to map parameters to genes, and what to do with hyperparameters such as the crossover method and the generation alternation method. On the other hand, there is a lot of work that needs to be done for each problem.
Therefore, although there are MoG-VQE\cite{2020arXiv200704424C} and E-VQE\cite{2019arXiv191009694R} that optimize quantum circuits by mapping their structures to genes, no attempt has been made to optimize them using real-coded GA (rcGA). In this paper, we will discuss how to use rcGA for optimization. We performed the benchmark of Powell, BFGS, MNelder-Mead, Bayesian, and rcGA as an optimizer of the VQE method.

In Chapter \ref{1}, the background and issues are presented. In Chapter \ref{2}, we propose a software circuit and a flowchart of the VQE method applying the GA method to find the minima of the energy states of the hydrogen molecule selected as the target molecule as the first step. In Chapter \ref{3}, we describe the detail of our calculation. In Chapter \ref{4}, we compare the accuracy of the GA method with that of other optimization methods for the ground state energy of molecular hydrogen. Chapter \ref{5}concludes our works.

\section{method}\label{2}
The VQE method and the genetic algorithm will be explained step by step. The VQE method and the genetic algorithm are explained step by step. The flow of finding the energy minima of the hydrogen molecule selected as the target molecule is explained using mathematical equations. A flowchart showing the software circuit and the calculation flow is also presented.
\subsection{VQE method}\label{2-2}
In NISQ devices, the VQE method is a well-known algorithm for deriving the minimum energy of a quantum state. This algorithm is a way to solve the energy minimization problem in a quantum system represented by the following quadratic quantized Hamiltonian: H

\begin{equation}
H = \sum_{j,k = 0}h_{jk}c_j^\dagger c_k + \sum_{j,k,l,m = 0}\langle jk \mid\mid lm \rangle c_j^\dagger c_k^\dagger c_l c_m. \label{molham}
\end{equation}

The energy eigenvalue $E$ of this Hamiltonian is

\begin{equation}
E=\langle \Phi \mid H \mid \Phi \rangle.\label{vqearchitype}
\end{equation}

The energy of the ground state is calculated by finding the state $\mid \Phi \rangle$ that minimizes this equation.
The state of the system is represented by a Slater determinant basis with occupied and unoccupied orbitals. For example, in the case of molecular hydrogen, there are two bonding orbitals and two antibonding orbitals. Hence, the ground state is represented as $\mid1100\rangle$, where 1 means the orbital is occupied and 0 means it is unoccupied, the two indices on the left correspond to bonding, and the two indices on the right correspond to antibonding orbitals. The two indices on the left correspond to couplings, and the two indices on the right correspond to anti-coupling orbits. The two indices on the left correspond to bonds, and the two indices on the right correspond to anti-bond orbitals. Matter tries to exist in a stable form with as little energy as possible. Comparing bonding orbitals and antibonding orbitals, bonding orbitals have lower energy and are more stable.
The bonding orbitals are represented by the STO-3G basis function system. All cluster terms and Hamiltonians are represented as Pauli matrices (operators) by Jordan-Wigner or Bravyi-Kitaev transformations\cite{doi:10.1021/acs.jctc.8b00450,openfermion}. The transformed Hamiltonian $H$ is

\begin{eqnarray}
H &=& f_01+f_1\sigma^z_0 +f_2\sigma^z_1 +f_3\sigma^z_2 +f_1\sigma^z_0\sigma^z_1 \\ \nonumber
&+& f_4\sigma^z_0\sigma^z_2 +f_5\sigma^z_1\sigma^z_3 \\ \nonumber
&+& f_6\sigma^x_0\sigma^z_1\sigma^x_2 +f_6\sigma^y_0\sigma^z_1\sigma^y_2 \\ \nonumber
&+& f_7\sigma^z_0\sigma^z_1\sigma^z_2 \\ \nonumber
&+&f_4\sigma^z_0\sigma^z_2\sigma^z_3 +f_3\sigma^z_1\sigma^z_2\sigma^z_3 \\ \nonumber
&+& f_6\sigma^x_0\sigma^z_1\sigma^x_2\sigma^z_3 +f_6\sigma^y_0\sigma^z_1\sigma^y_2\sigma^z_3 +f_7\sigma^z_0\sigma^z_1\sigma^z_2\sigma ^z_3. \\
\label{bkham}
\end{eqnarray}

In quantum chemical calculations, the ground state is calculated first, and then the excited states are calculated.

In the case of molecular hydrogen, it is possible to calculate the ground state from the Hamiltonian alone, but in general, this is not enough to calculate the wave function of the excited state.
In general, however, this is not enough to calculate the wavefunction of the excited states. Therefore, we need to prepare a cluster term $T$ that causes transitions between all allowed states.
This is mainly referred to as Unitary-Coupled-Cluster (UCC).
In this paper, we apply the UCCSD, which deals only with two-electron excitations. The cluster term $T$ is given by

\begin{equation}
T=\sum_{j\in occu.,k\in vac.}c_j^\dagger c_k + \sum_{j,k\in occu.,l,m\in vac.}c_j^\dagger c_k^\dagger c_l c_m.
\label{clus}
\end{equation}

\begin{figure}[h].
\includegraphics[scale=0.7]{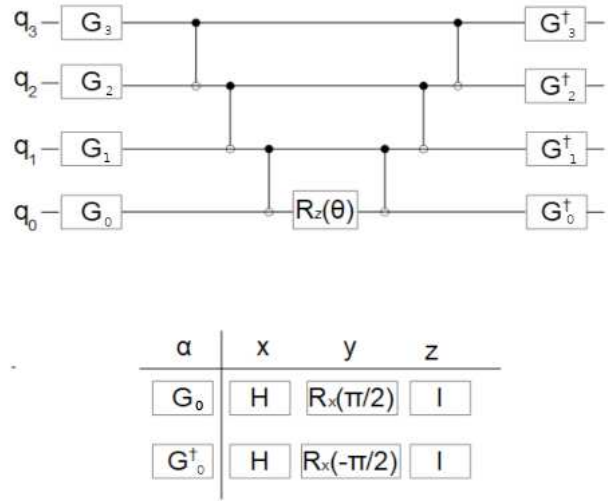}
\caption{The quantum circuit that is used for the VQE method that performs $exp(-i\theta\sigma_0^\alpha\sigma_1^\beta\sigma_2^\gamma\sigma_3^\delta)$. $G_j$ gate is $H$, $R_x(\pi/2)$ or $I$ gate depend on what pauli gate is operated for $q_j$.}\label{vqeexp}
\end{figure}

The result is multiplied by the quantum state in the form $exp(i(T-T^\dagger))$ to treat the excited state.
The Suzuki-Trotter decomposition decomposes the Hamiltonian and cluster terms into products of exp \cite{McClean_2016}.
To make all the coefficients of the Hamiltonian sufficiently small, increase the depth (number of iterations) d of the circuit.
This way, there is no need to add variables to the Hamiltonian.
However, this would require an infinite amount of time for quantum computation, so it is desirable to keep the circuit depth to about 2.
To do this, all the terms are transformed by adding variables, and then the variables are optimized.
All terms transformed to Pauli operators are converted to quantum circuits in the form $exp(-i\theta_kP_jt)$.
Here, $\theta_k$ is the optimization variable in the $k$th term, $P_j$ is the $j^{th}$ Pauli operator combination belonging to $k$, and $t$ is the coefficient of the operation. Thus, we are ready to make the excited state. All terms can be expressed as a product of time-promoting operations, and a quantum computer can handle them as gate operations. The circuit would look like Fig. \ref{vqeexp}.

This circuit operates on terms consisting of multiple Pauli operators in a four-qubit system. The $q_0-q_3$ represent the first through fourth qubits, $q_0,q_1$ represent the coupled orbits, and $q_2,q_3$ represent the anti-coupled orbits. The $R_z(\theta)$ exists to multiply the coefficients, and the $G_j$ and $G_j^\dagger$ are determined by which the Pauli operator is multiplied to the $j$-th bit.
For example, when multiplying $exp(-i\theta_k\sigma_0^x\sigma_1^y\sigma_2^z)$, $G_0$ is the $H$ gate, $G_1$ is $R_x(\pi/2)$, and $G_2$ is the $I$ gate (unit gate). Since there is no Pauli operator acting on $q_3$, there is no CNOT between $G_3$ and $q_2,q_3$. Also, when multiplying by $exp(-i\theta_k\sigma_0^x\sigma_2^y\sigma_3^z)$, the CNOT gate between $q_0$ and $q_1$ replaces the gate between $q_0$ and $q_2$, the CNOT gate between $q_1$ and $q_2$ and $G_2$ disappear, and $G_3$ and $G_3$ are no longer in use. The CNOT gate between $q_1$ and $q_2$ and $G_2$ disappear, and $G_3$ takes the place of $G_2$. Everything else is the same. In this way, a trial state is created for the excited state by applying the cluster term and the Hamiltonian to the initial state, and the parameters are changed according to the value of the evaluation function to find the minimum and its energy eigenstate $\mid \Phi_i \rangle$. The optimization evaluation function $F_i(\bm{\theta})$ for the $i^{th}$ state is

\begin{eqnarray}
F_i(\bm{\theta}) &=& \langle \Phi_{ini}\mid UHU^\dagger \mid \Phi_{ini}\rangle \\ \label{F}
&+& E_i^{def} + E_i^{const} \nonumber
\end{eqnarray}

\begin{equation}
U= \Pi_{j,k} exp( i \theta_k P_j t_j).\label{U}
\end{equation}

Then, $\mid\Phi_{ini}\rangle$ is the ground state of the Bravi-Kitaev transformed molecular hydrogen orbital, $\mid1000\rangle$, and $E_i^{def}$ and $E_i^{const}$ are the deflation and constraint conditions for the $i^{th}$ state, respectively. The last two terms go to zero when the energy reaches the optimal solution. These terms are necessary to derive the excited states. The method is the Variational Quantum Deflation (VQD) method. This method is one way to derive the energy of an excited state. This is done by adding the projection operator of the state with lower energy levels to the evaluation function. Its form $E$ is

\begin{equation}
E=\langle\Phi_i\mid(H+A\sum_{j<i}\mid\Phi_j \rangle \langle\Phi_j\mid)\mid\Phi_i\rangle\label{def}
\end{equation}

The result is where $\mid \Phi_i\rangle, \mid \Phi_j\rangle $ denote the $i,j$th state, respectively.

The product between the $i^{th}$ state and the $j^{th}$ state is derived by the SWAP test algorithm \cite{2013PhRvA..87e2330G}, where $A$ is the weight factor. The deflation term $E_i^{def}$ for the $i^{th}$ state is

\begin{eqnarray}
E_{i}^{def}&=&((af+b(1-f)))\\ \nonumber
&\times& (\sum_{j<i}(exp(r-0.25r_d)+1)^{-1}) \\
&\times&\mid\langle\Phi_j\mid\Phi_i\rangle\mid{^2}\\ \nonumber
&+&(1-(exp(r-0.25r_d)+1)^{-1}) \\ \nonumber
&\times & f(\mid\langle\Phi_j\mid\Phi_i\rangle\mid{^2})).\label{defmisc}
\end{eqnarray}

$a$, $b$, and $f=(exp(\alpha(r-r_d))+1)^{-1}$ are functions on two constants and $r$ for Fermi-Dirac, respectively. Furthermore, $r_d$ is a constant bond length in molecular hydrogen. $f(\mid\langle\Phi_j\mid\Phi_i\rangle\mid{^2})$ is a quadratic function of the absolute value of the product between the $i$ and $j^{th}$states to accurately derive the degenerate state. For molecular hydrogen, $a=1.0$, $\alpha=100$ and $f(\mid\langle\Phi_j\mid\Phi_i\rangle\mid{^2})=(1+2(\sqrt{5}+1))r^4/r_d^4E_p(r)/4\mid\langle\Phi_j \mid\Phi_i\rangle\mid{^4}+2(\sqrt{5}+1)r^4/r_d^4E_p(r)/4\mid\langle\Phi_j\mid\Phi_i\rangle\mid{^2}$. where $E_p(r)$ is the energy of one lower state of molecular hydrogen at bond length $r$. The constraint calculation can be performed by adding a constraint to the equation \ref{def}. Thus, the constraint term in the $i^{th}$ state is

\begin{equation}
E_{i}^{const}=\sum_{j = 0}^{num of const.}\langle \Phi_i \mid(U_j - U_{j}^{const})\mid \Phi_i \rangle \label{const}
\end{equation}

, where $U_j$ denotes the physical quantity to be bound and $U_{j}^{const}$ denotes the target value.
Here, $U_j$ is assumed to be the square of the spin variable ${\bf S}$ and the spin $\sum_{i}s_i^z$.

\begin{figure*}[h].
\includegraphics[scale=0.9]{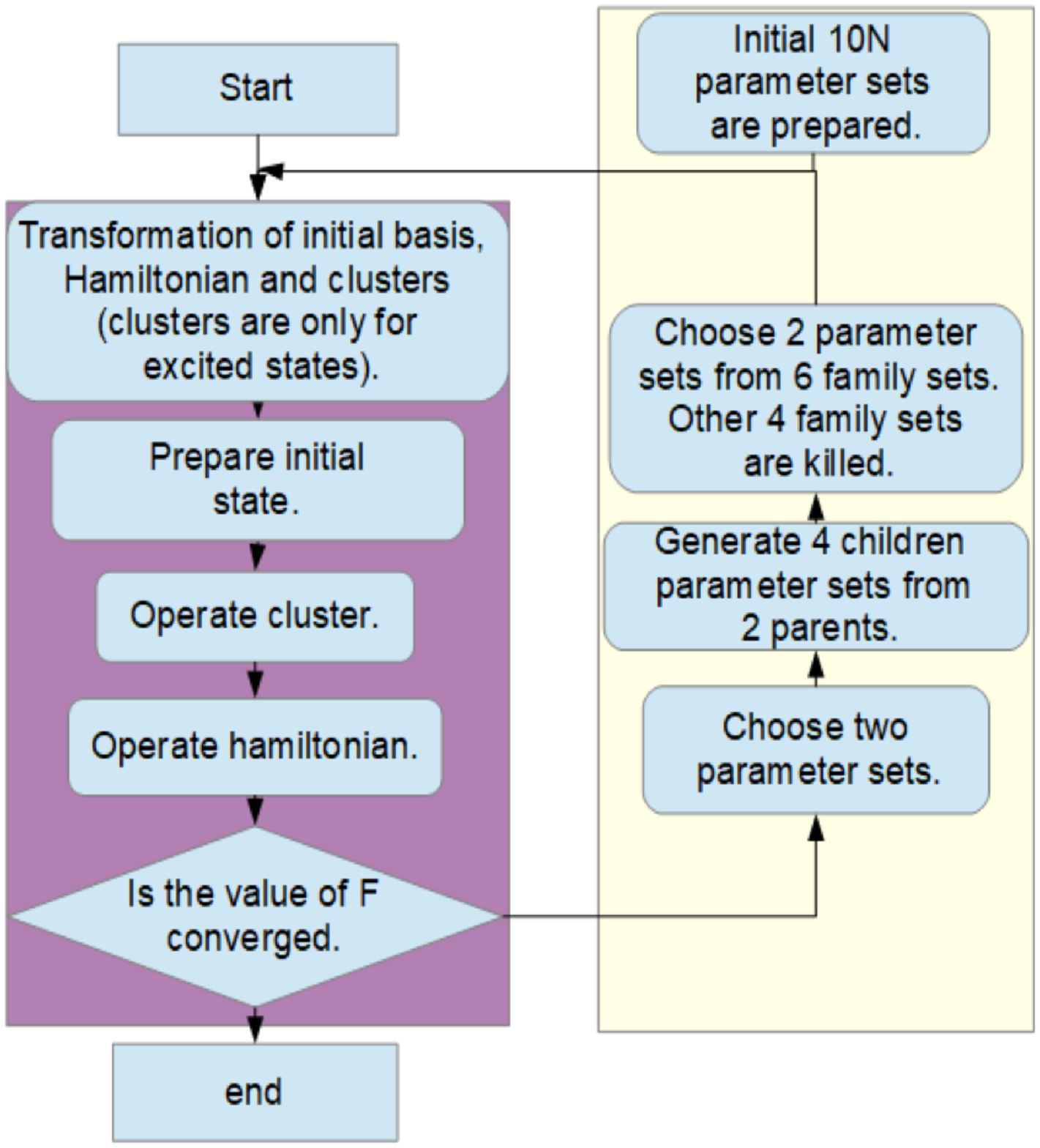}
\newline
\newline
\caption{The flowchart of the VQE method using rcGA for its optimizer. The purple rectangle indicates inside processes are performed by quantum computers and Purple rectangle indicates inside processes are performed by quantum computers and the yellow rectangle indicates inside processes are performed by classical computers, respectively. Calculation of eigenenergies is quantum computers' Calculation of eigenenergies is a quantum computers' job and optimizing parameter sets $\bm{\theta}$ by rcGA is a classical computers' job, respectively. In case we use other optimizers, the processes in the yellow rectangle are performed by the other optimizers.} \label{vqediag}
\end{figure*}

For optimization, we use the value obtained by computing the optimization evaluation function including energy, condition, and deflation terms in terms of the parameters.

In the parameter landscape of the VQE method, there are many pockets, such as the Rastregin function, as the level of the excited state increases, and there is a minimum or local minimum. Therefore, Newtonian optimization is prone to be captured by them.

\subsection{Genetic Algorithm used for VQE}\label{2-3}
Genetic Algorithm (GA) is one of the algorithms for finding the global optimal solution to a combinatorial optimization problem. This algorithm can be applied to real continuous variables. It is called real-coded GA(rcGA). Among them, the method combining Just-Generation-Gap(JGG)\cite{Kita1999,Akimoto2007}and Real-Coded Ensemble Crossover(REX)\cite{TAKAHASHI2013239}can effectively find the global optimal solution in a relatively short time. The combination of Real-Coded Ensemble Crossover (REX) and JGG\cite{TAKAHASHI2013239} is a powerful method \cite{2009N-EC_Invited} because it can effectively output the global optimal solution in a relatively short time. REX is the crossover method, which calculates the parameters of the children based on the center of gravity and variance of the parameters of the parents, and is performed according to the following equation.

\begin{equation}
\bm{\theta}^{(g,i)} = \bar{\bm{\theta}}^{(g-1)} + \sum_j^{N_p}\xi_j(\bm{\theta}^{(g-1,j)} - \bar{\bm{\theta}}^{(g-1)})\label{rex}
\end{equation}

where $N$ is the number of parameters to be optimized and $\xi_j$ is a random variable subject to variance, corresponding to mutation.
It is a uniform random variable with its average 0 and standard deviation $0.9/\sqrt{N_p}$ for the number of parent parameters $N_p$.
$\bm{\theta}^{(g-1,j)}$ denotes the parameter vector of the $j$-th parent in generation $g-1$ of parameter vector $\bm{\theta}$, and $\bar{\bm{\theta}}^{(g-1)}$ denotes the parameter vector $\bm{\theta}$ in generation $g-1$ of parameter vector $\bm{\theta}$. in the $g-1$ generation. JGG is a generational alternation method and is performed as follows. First, for a parameter $N$, $10N$ individuals are randomly generated in the first generation. The initial values are generated according to the following equation.

\begin{equation}
\bm{\theta}_j^{(0, i)} = (\bm{\theta}_j^{UB}-\bm{\theta}_j^{LB})f_{ini} + \bm{\theta}_j^{LB}
\end{equation}

where $\bm{\theta}_j^{UB},\bm{\theta}_j^{LB}$ are the upper and lower limits of the $j^{th}$ parameter, respectively.
Also, $f_{ini}$ is the initialization function.
In the second generation, $2$ parents are randomly selected from the population, and $4$ children are generated from them and form the family by them.
From there, the evaluation function $F_n(\bm{\theta}^{(g, i)})$ is computed, and the $2$ individuals selected in order of closeness to the desired evaluation value are returned to the population. This is one generation so far.
Thereafter, the generation is updated iteratively. The number of children and parents is determined to save the calculation time keeping the accuracy.
Normally, the mean gap should be the convergence condition, but in this case, since the target evaluation value of the evaluation function is set to negative infinity, the standard deviation of each variable in $F_n(\bm{\theta}^{(g, i)})$ is set as the convergence condition. The convergence condition is

\begin{eqnarray}
\sigma_k &=& \sum_{j}^{10N}(\bm{\theta}^{(g, j)}_k - \bar{\bm{\theta}^{(g)}_k})^2/10N \\
10^{-16} &>& Max(\sigma_k/(\bm{\theta}^{UB}_k - \bm{\theta}^{LB}_k)).\label{conv}
\end{eqnarray}

Here, $\sigma_k$ is the standard deviation in the $k$th variable. In this case, the process of creating children and calculating their evaluation functions can be done in parallel, as long as the number of qubits and computational resources of the quantum computer allows, by submitting the calculation of the evaluation function for each child to an individual quantum computer as a job, which is how it is done in this paper.
This is a unique advantage of rcGAs, which cannot be found in other classical optimization algorithms for quantum computation.

In the calculation of the excited states, the deflation term is calculated, so if all variables are optimized, the number of individuals will be 400, and the calculation time will be too long.
Therefore, if you don't want to, you can optimize only the cluster variables in the excited state calculation, and use the ground state variables for the Hamiltonian.

Using rcGA to calculate the eigenenergies of molecular hydrogen in the ground state, the method appears to be sufficiently resistant to convergence to local minimums and sufficiently accurate to exceed chemical accuracy.
As mentioned above, today's quantum computers are called NISQ computers, and it is better to let them handle only the operations that they are good at, and let the classical computers handle the other calculations.
Therefore, as shown in Fig. \ref{vqediag}, the classical computer is in charge of the rcGA optimization, while the quantum computer is in charge of the process of creating quantum states and calculating their energy eigenvalues.
Following this flowchart, we compared and examined the results of the ground and excited states of the VQE method when the classical optimization part is rcGA and when it is not.
\newpage

\section{Calculation on hydrogen molecules}\label{3}
The depth of the circuit for calculating the cluster and Hamiltonian are both 2. Therefore, the number of variables for the cluster is 10, and the number of variables for the Hamiltonian is 30.

The VQE method was performed on blueqat SDK\cite{blueqat}, a quantum computer simulator.

The VQE method was performed on a quantum computer simulator, blueqat SDK\cite{blueqat}, which is based on Newton's method, vector method (a) Powell's method, (b) CG method, (c) Nelder-Mead method, (d) BFGS method, and (e) Bayesian method to calculate the global optimal solution. (e) Bayesian method and (f) rcGA method. All data are averaged over five runs. The number of shots on the simulation of quantum calculation is infinity (using the state vector as a result of quantum calculations).
The data are all averaged over five times, each in the range $r\in[0.1,2.5]$ in increments of 0.1.

The optimization methods used in (a)-(d) are those of scipy.

The number of iterations for (a) Powell's method and (c) Nelder-Mead's method is 2000, that for (b) CG method is 20, and that for (d) BFGS method is 50. The number of iterations for (b) CG method is 20, and that for (d) BFGS method is 50, to keep the computation time in the same time range for (a)-(d).
Gaussian-Process-Regression was applied to the Bayesian method, and up to 100 iterations of the prior probability distribution calculation with 30 dimensions in the ground state and 10 dimensions in the excited state were performed with GPyOpt\cite{GPyOpt}.
The unit of energy was Hartree.

On the other hand, to take advantage of the characteristics of the rcGA algorithm, the Hamiltonian of the ground state was also used for the excited states (triplet, singlet, and doubly excited states) at each interval. Furthermore, the calculation was divided into two cases: one where only the variables of the Hamiltonian are determined (f), and one where the variables of the Hamiltonian are not determined. The number of alternations is 3000, 10000, 10000, and 15000 generations for the ground state, triplet state, singlet state, and two-electron excited state, respectively. The number of individuals is 300, 100, 100, 100, and 100, respectively, and since all individuals are given equal opportunity to move, the number of times each individual is updated is 100, 100, 100, and 150 times for each state.
The variables in the initial population are a mixture of the beta distribution with a$=$b$=$0.99 and the uniform distribution with a ratio of 1000:1.
\section{Result and Discussion}\label{4}
\begin{figure*}[h]
\includegraphics[scale = 0.9]{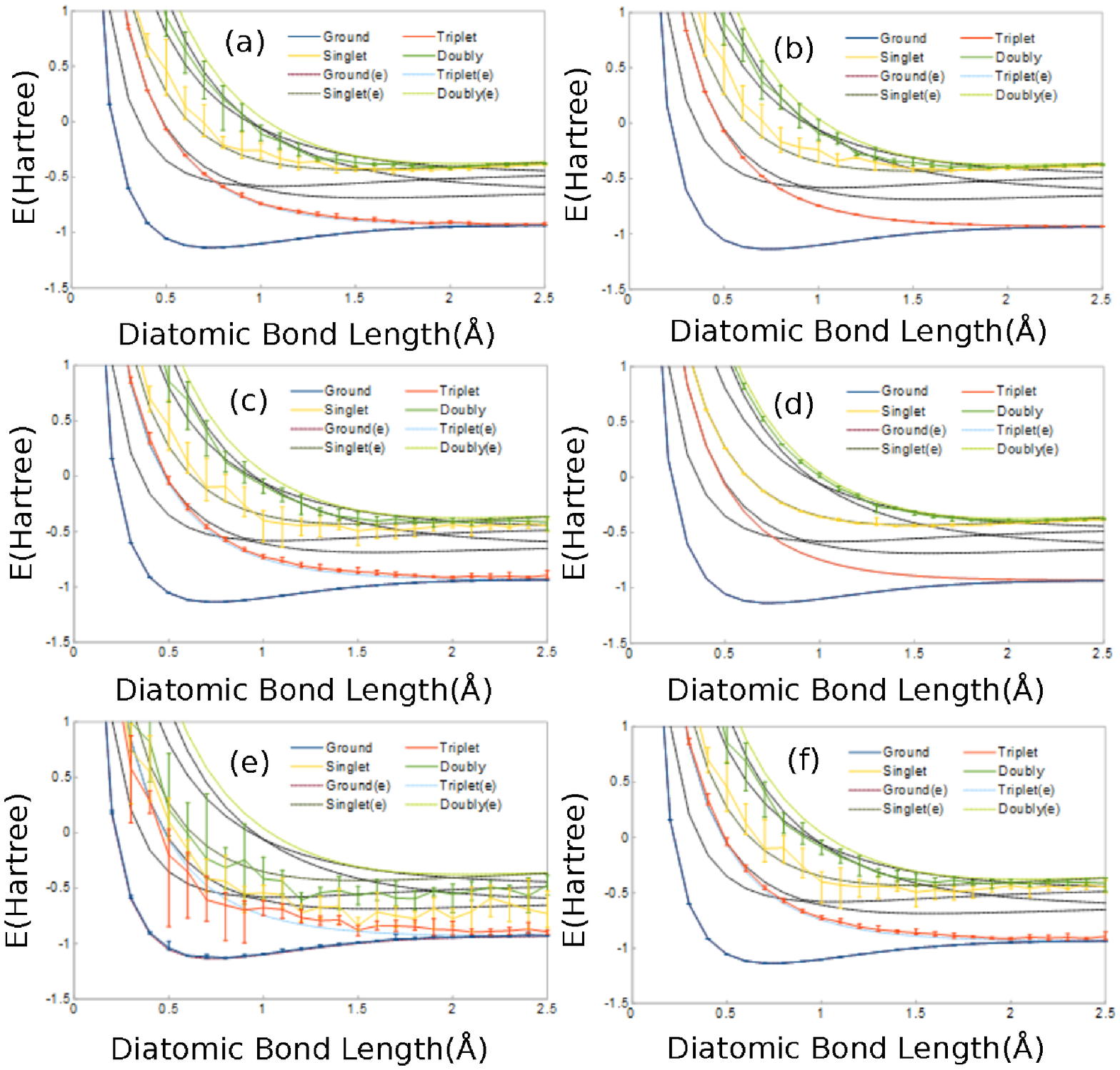}
\newline

\caption{The diatomic bond length v.s. the energy levels of ground, triplet, singlet, and doubly excited states, calculated by (a)Powell, (b) Conjugate-Gradient, (c)Nelder-Mead, (d)BFGS, (e)Bayesian optimization, and (f)rcGA that the variables of hamiltonian are fixed in those of ground state, respectively. All points are sampled in 0.1 pitch from 0.1 to 2.5 for diatomic bond length. The lines that have (e) in their suffix are exact values calculated by the Full-CI method. All data are the averages on five samples. Black lines indicate the doublet states that are local minimums.}\label{cmp}
\end{figure*}

\begin{figure*}[h]
\includegraphics[scale = 0.8]{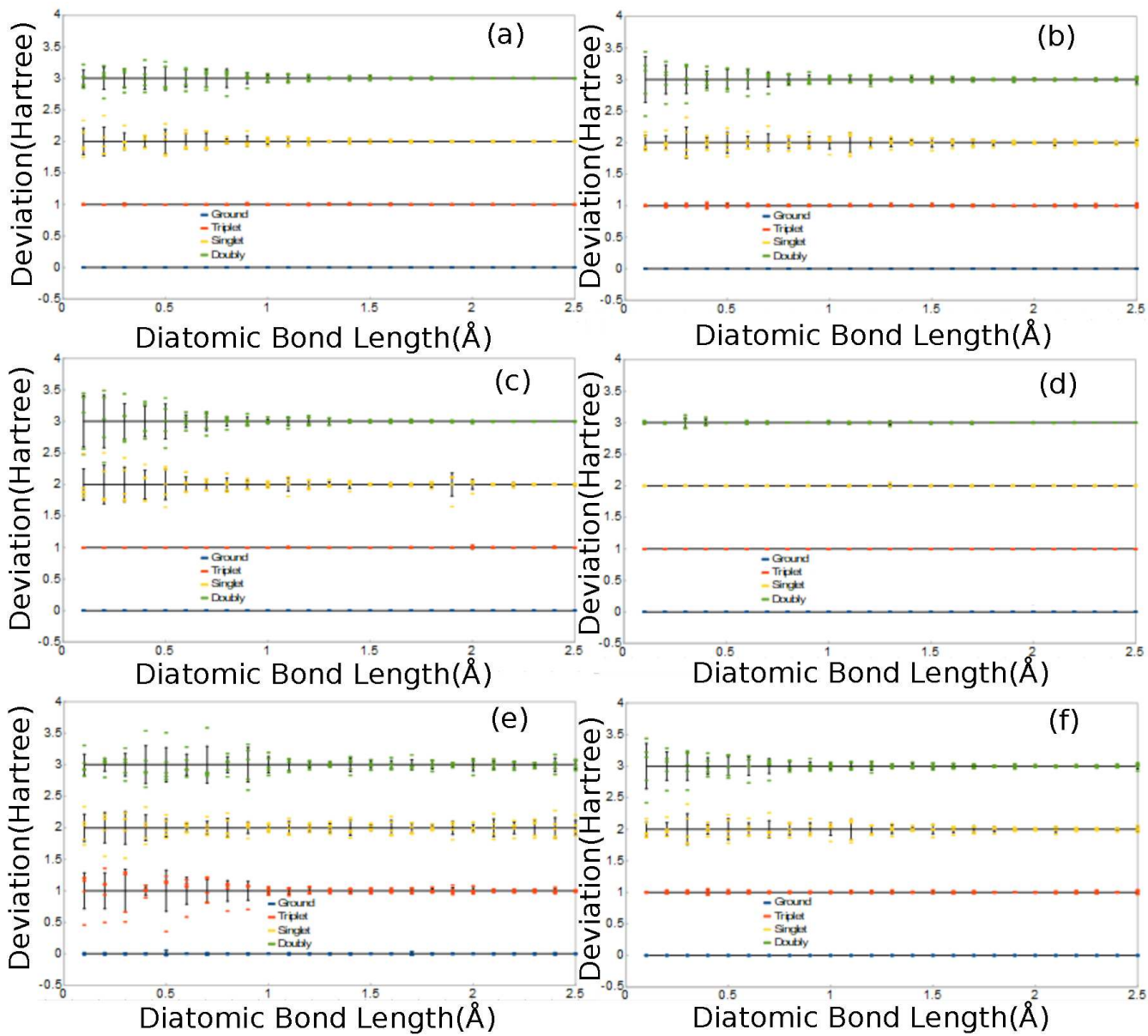}
\newline

\caption{The diatomic bond length v.s. the deviation of each calculated energy from its average of ground, triplet, singlet, and doubly excited states, respectively calculated by (a)Powell, (b) Conjugate-Gradient, All points are sampled in 0.1 pitch from 0.1 to 2.5 for diatomic bond length. (c)Nelder-Mead, (d)BFGS, (e)Bayesian optimization, and (f)rcGA that the variables of hamiltonian are fixed in those of ground state, respectively.
Black lines indicate the average of ground, triplet, singlet, and doubly excited states, respectively. Ground, triplet, singlet, and doubly excited state are Deviation(Hartree)=0, 1, 2, and 3, respectively.
Black error bars indicate the standard deviations of each sampled point.}\label{dev}
\end{figure*}

The calculated energies of the ground state, triplet state, singlet state, and two-electron excited state with respect to the interatomic distance between hydrogen atoms are shown in Fig. \ref{cmp}, and their deviations and standard deviations for five calculations for each interatomic distance $r$ is shown in Fig. \ref{dev}.
It can be seen that the energy levels and their error bars in (a)-(d) for the ground state and triplet state are almost identical to the exact solution. However, there is a clear difference in the singlet and doubly excited states. The averages of (a) Powell's method, (b) CG method, and (c) Nelder-Mead method are often inaccurate due to the attraction of local minimums. However, in the (d) BFGS method, the exact value, the calculated average value, and the error range of the singlet state are almost identical. For the doubly excited states, the mean value was a little below the exact solution, and the error range was almost the same at all points.

According to the deviation of the calculation results, the calculation results for the singlet state and the two-electron excited state varied greatly compared to the lower two levels as in Fig. \ref{cmp}. For (a) Powell's method, the variation was the second-lowest among (a)-(d), after (d) BFGS method. The Nelder-Mead method (c) had the largest variation among (a)-(d), and the BFGS method (d) had the smallest variation.

The results of the Bayesian optimization in (e) show that there are deviations from the exact solution not only for the singlet and doubly excited states but even for the triplet state.
The error bars are also large, and the lower bound of the error bars for the singlet and doubly excited states extends to the doublet local minimum below the exact solution for the singlet state.

The deviations of the calculation results are the most scattered among (a)-(f).

The results for rcGA in (f) show that even the triplet state is a little far from the exact solution in the region $r>0.9$.
The results for the singlet and doubly excited states are far from the exact solution, and the lower limit of the error bars is below the exact solution for the singlet state and overlaps with the local minimum.

According to the deviation of the calculation results, the standard deviation at $r\leq 1.0$ in the singlet state is slightly smaller than that of (b) CG method after (d) BFGS.
In the two-electron excited state, the standard deviation in the region of $r\leq 0.5$ is the closest to the exact solution.

\begin{figure*}[h]
\includegraphics[scale = 0.85]{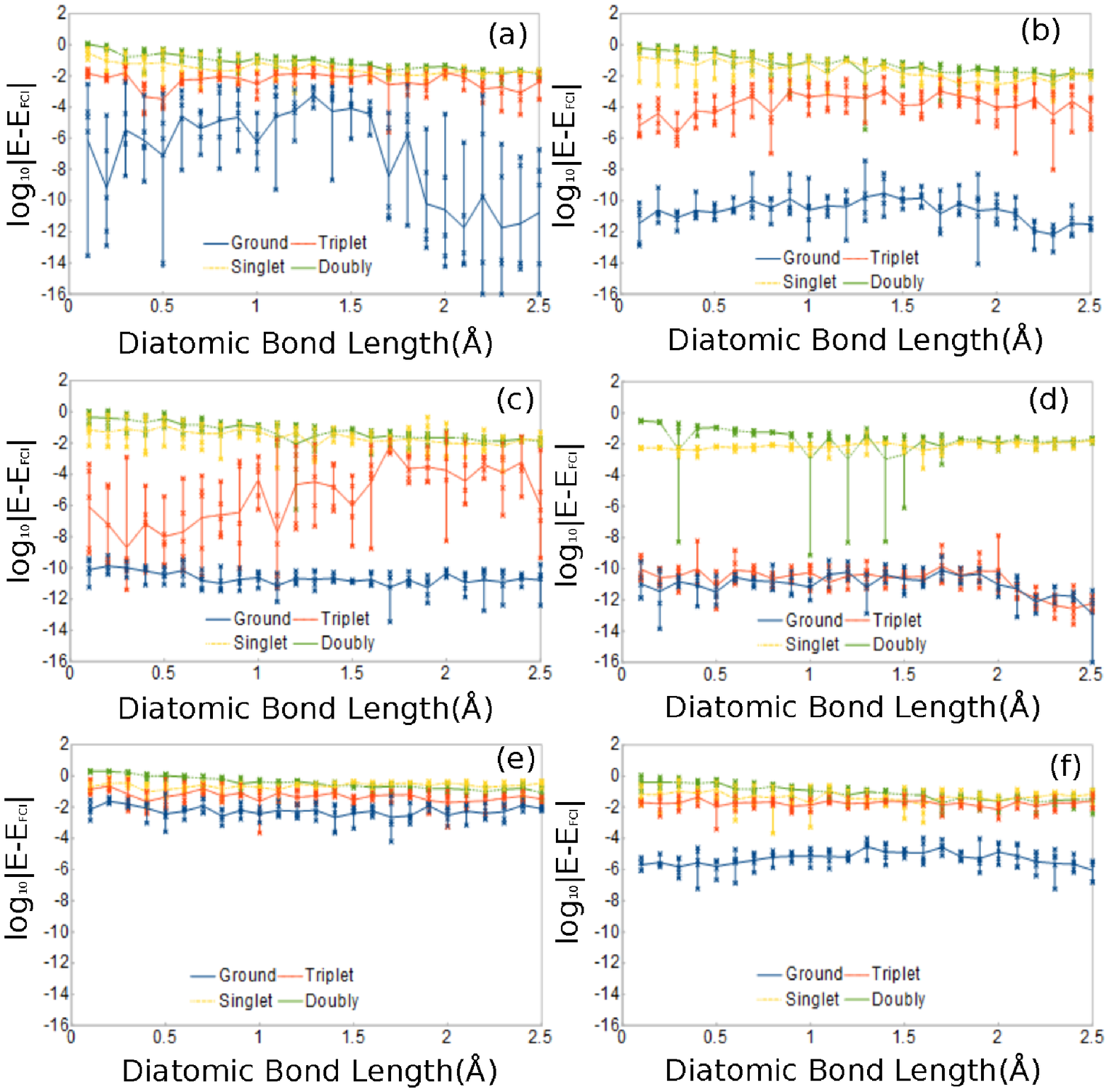}
\newline

\caption{The diatomic bond length v.s. the ordinal log of the difference between the average of calculated energies and the exact value calculated by STO-3G classically (log error) of the energy levels of ground, triplet, singlet, and doubly excited states, respectively calculated by (a)Powell, (b) Conjugate-Gradient, (c)Nelder-Mead, (d)BFGS, (e)Bayesian optimization, and (f)rcGA that the variables of hamiltonian are fixed in those of ground state, respectively. All data are log errors of the averages sampled five times. Error bars indicate the realm between the minimum log error of one sample and the maximum log error of another sample. Each point indicates the log error of each sampled calculated energy.}\label{cmpr}
\end{figure*}

The logarithmic error and the logarithmic error of the difference from the exact solution in the STO-3G basis of the averaged energy values for the distances between hydrogen atoms in the ground state, triplet state, singlet state, and two-electron excited state are shown in Fig. \ref{cmpr}.
(a) The logarithmic error in Powell has a large range in all levels, and the lower limit of the error bars in the singlet and doubly excited states is less than $-2$, while the upper limit is greater than $0$.
Even in the ground state, there were many points where the maximum logarithmic error was less than the chemical accuracy in the whole range.
(b) The logarithmic error in the CG method is in the range of -10 to -12 in the ground state, but there are several points where the error bars become wide. The values for the singlet and doubly excited states are mostly around the minimum of -2, although there are a couple of points that exceed the chemical accuracy.
(c) The logarithmic error of the Nelder-Mead method is similar to that of (b) for the ground state, but the error bars are noticeably wider. The logarithmic error in the singlet and doubly excited states falls to a minimum value of about -2.
(d) The logarithmic error of the BFGS method for the ground state and triplet state is about -10 to -12, and the error bar is the smallest among (a)-(d). For the singlet and doubly excited states, the average error falls between -2 and -3, but the error bars are also small, and the logarithmic error for the doubly excited states does not exceed the chemical accuracy except for four values below -8 at the minimum. Although the results of this method are highly accurate and stable, the number of iterations sets a limit to the accuracy of the above two levels, and it is very rare to achieve higher accuracy.

The logarithmic error in the Bayesian optimization in (e) was so large that it could not exceed the chemical accuracy even in the ground state, although the error bars were small, and the logarithmic error was noticeably large in all states. This may be because the Gaussian-process-regression Gaussian probability distribution matrix does not identify pockets of global minima in the parameter space.

Looking at the deviations, even for the triplet state, there were many points where only one out of five calculations resulted in a highly accurate value. For the singlet state and the two-electron excited state, as shown in Fig. \ref{cmp}, there were almost no points for which a high-precision value was calculated.

On the other hand, the error bars for rcGA in (f) were only in the range of 0.5 to 2.5 for all levels. The accuracy of the ground state greatly exceeds the chemical accuracy, including error bars, over the entire range of inter-atomic distances. Although, the accuracies on excited states are low. The global minimums of excited states are on the bottoms of valleys of parameter landscape. Hence, the next generations must be generated taking into account them. rcGA never uses the deviation and Hessian matrices for optimization, thus, the results are trapped by local minimums or couldn't reach global minimums. The same tendency appears in the results of (a)Powell's method.
Looking at the deviation of the ground state, the minimum of the error bars was reached or approached more than twice in a relatively large number of points in (a)-(f).
This means that by distributing individuals uniformly throughout the parameter space and moving all of them evenly, the final distribution of individuals can be generated by changing generations a sufficient number of times, as long as the range that can be moved by the child generation function eq. \ref{rex} is not greater than the pocket radius of the global minimum in the parameter space.

The logarithmic errors for the singlet and doubly excited states are about -2 on average, but there are five error bars for both levels that exceed the chemical precision.

\newpage\newpage

\newpage
\newpage
\newpage\newpage

In the rcGA, we studied the initial population distribution. From now on, based on the result of Fig.\ref{cmpr}(f), the calculation is performed only once because the error range is considered to be small regardless of the initial population distribution.
The number of generation changes is set to 3000, 15000, 15000, and 25000 for the ground state, triplet state, singlet state, and two-electron excited state, respectively.

\begin{figure}
\includegraphics[scale=0.2]{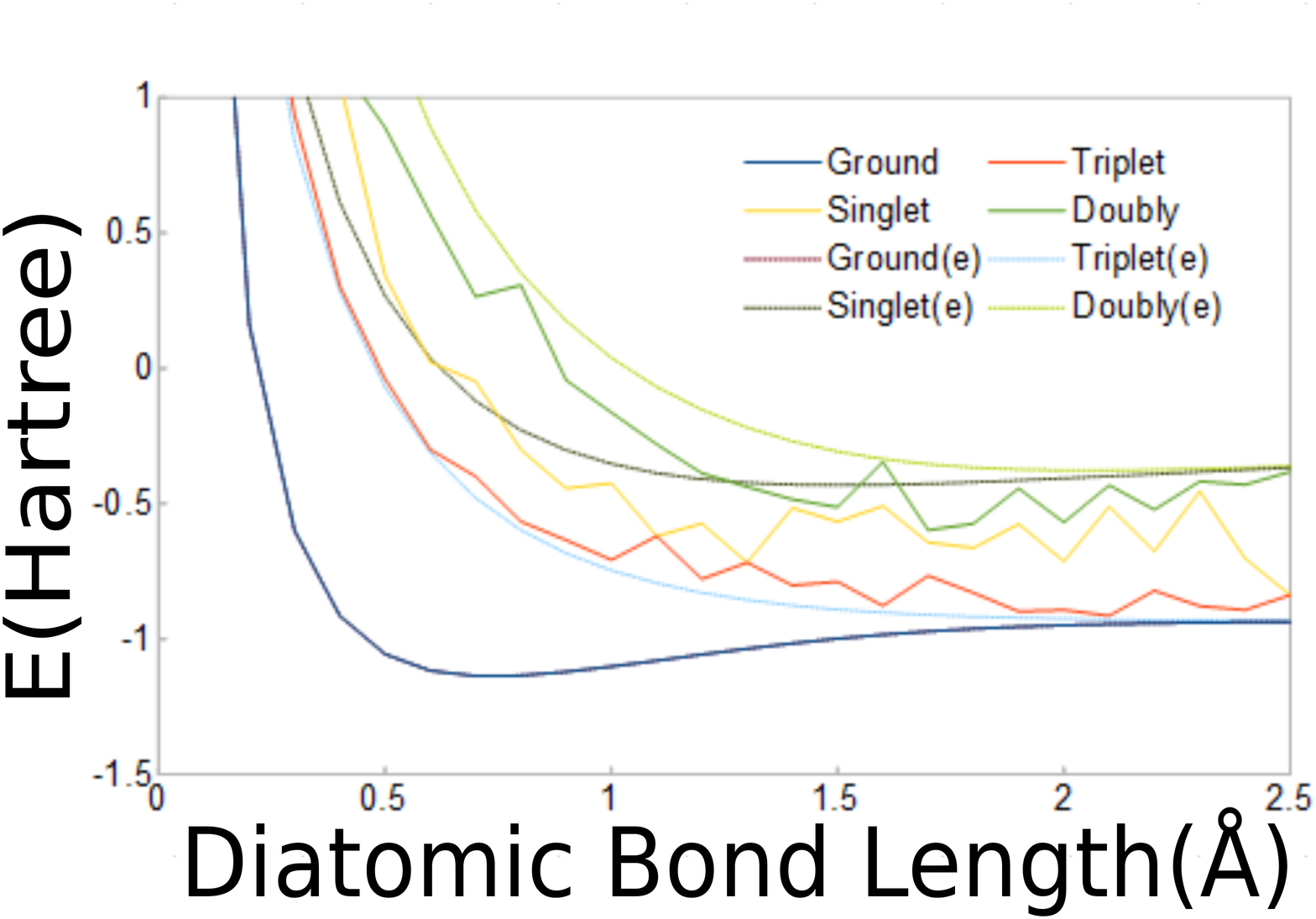}
\newline
\caption{The diatomic bond length v.s. the energy levels of ground, triplet, singlet, and doubly excited states, respectively calculated by rcGA that initial pool is arranged in the shape of Poisson distribution with $\lambda = 1$. The lines that have (e) are exact values calculated by Full-CI.}\label{rp}
\end{figure}

\begin{figure}
\includegraphics[scale = 0.2]{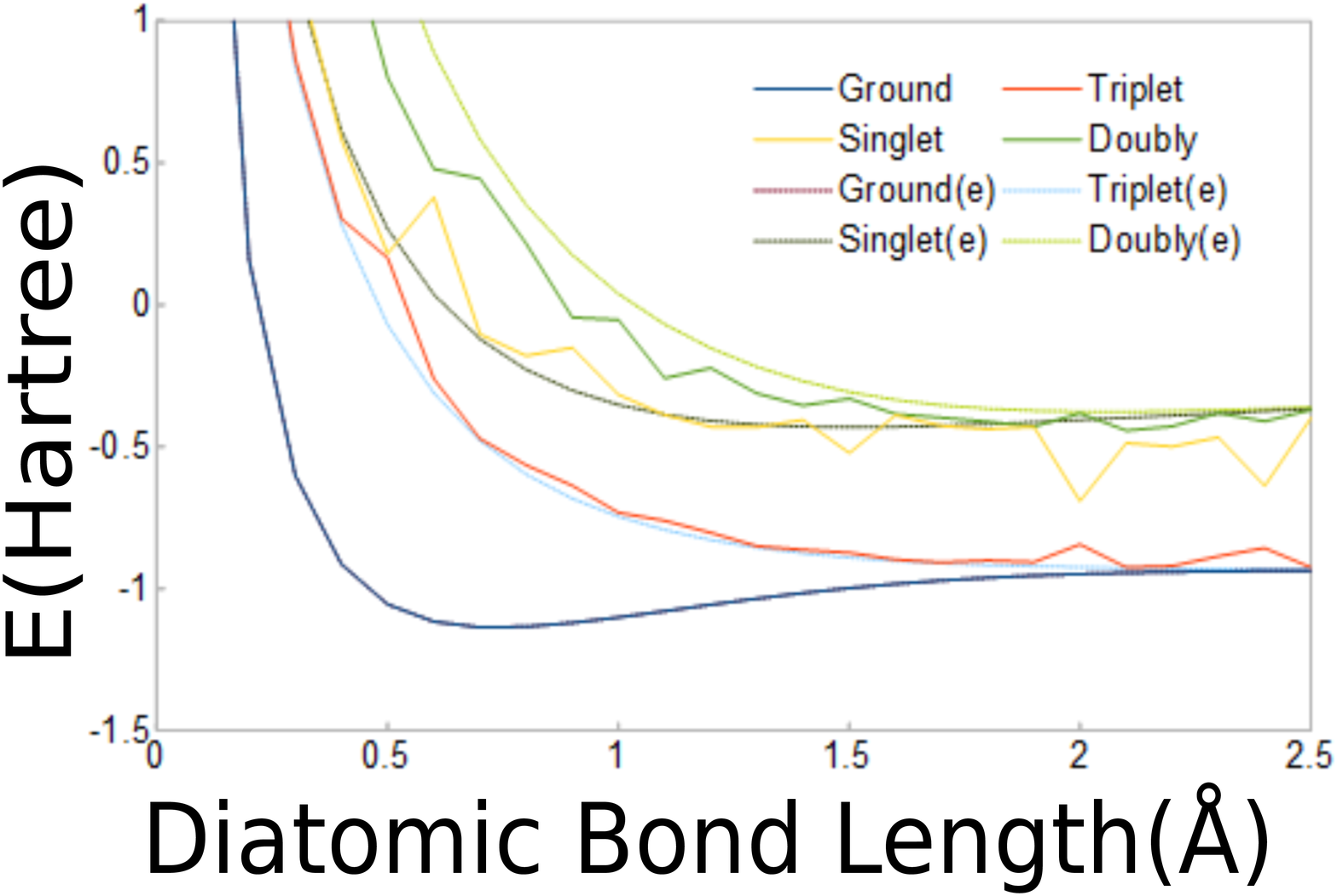}
\newline
\caption{The diatomic bond length v.s. the energy levels of ground, triplet, singlet, and doubly excited states, respectively calculated by rcGA that initial pool is arranged in the shape of Beta distribution with $a=b=0.25$. The lines that have (e) are exact values calculated by Full-CI.}\label{rb}
\end{figure}

Fig. \ref{rp} shows the results for the initial population following the Poisson distribution with $\lambda$$=$1.
Compared to Fig.\ref{cmp}, the results deviate significantly from the exact solution except for the ground state.
Fig.\ref{rb} shows the results for the initial population according to the beta distribution with a$=$b$=$0.25.
First, the computed potential surface is shown. For the Poisson distribution, after r$>$1.9, the singlet and doubly excited states depart from the exact solution beyond the error range in Fig.\ref{cmp}(f).
This is probably because the distribution is biased and the calculation starts from the state where the individuals are concentrated in the center of the parameter space.
On the other hand, in the $\beta$-distribution, convergence to the local minimum is suppressed, and the values of the singlet state and the two-electron excited state also deviate from the exact solution beyond the error range compared to Fig.\ref{cmp}(f), where the initial values of the populations are determined to be almost uniform.
However, the overall accuracy was better than the Poisson distribution. This is thought to be because the initial individual value is a$=$b, which is close to a uniform distribution.

\begin{figure}
\includegraphics[scale = 0.2]{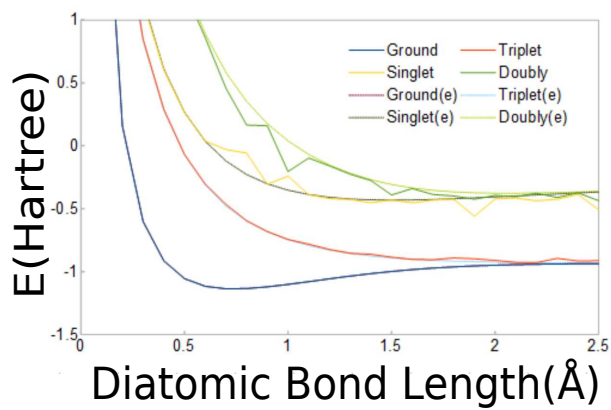}
\newline
\caption{The diatomic bond length v.s. the energy levels of ground, triplet, singlet, and doubly excited states, respectively calculated by rcGA that optimizes all variables at the same time. The lines that have (e) are exact values calculated by Full-CI.}\label{gafH}
\end{figure}

\begin{figure}
\includegraphics[scale = 0.2]{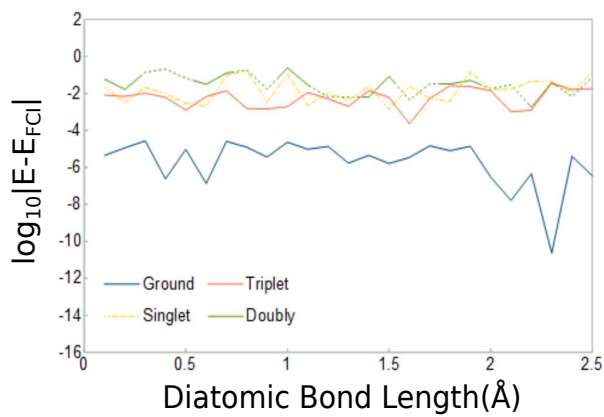}
\newline
\caption{The diatomic bond length v.s. the accuracies of calculation of energy levels of ground, triplet, singlet, and doubly excited states, respectively calculated by rcGA that optimize all variables in the same time. Each data is sampled for every 0.1 points from $r = 0$ to $r = 2.5$.}\label{gafaH}
\end{figure}

It was found that fixing the variables of the Hamiltonian at the values of the ground state has a limitation in terms of the improvement of accuracy.
The Hamiltonian is calculated not only for the ground state but also for the triplet, singlet, and doubly excited states, and the optimized results are shown in Fig. \ref{gafH}.
The number of generation changes was set to 30000, 40000, 120000, and 120000 generations for the ground state, triplet state, singlet state, and two-electron excited state, respectively.
The accuracy of the rcGA is improved by keeping the inter-individual spacing close to equal, as in Fig. \ref{rp} and \ref{rb}, as in Fig. \ref{cmp}(f).
The logarithmic error of the energy calculated by the rcGA that simultaneously optimizes all variables for the interatomic distances v.s. ground state, triplet state, singlet state, and two-electron excited state in molecular hydrogen is shown in Fig. \ref{gafaH}.
For some values, the rcGA converges to the local minimum, and for r$>$1.8, the accuracy is greatly improved, although the weakness to degeneracy is noticeable. The accuracies of these results never surpass that of (d) the BFGS method. The method for crossover and offspring must be improved taking into deviations and Hessians account.

\clearpage
\newpage
\section{Concluding Remaerks}\label{5}
In this paper, we revealed that BFGS is the most accurate optimizer for the VQE method in case the cluster is UCCSD. rcGA has low accuracy same as Powell's method. It is because rcGA optimizes each individual without derivatives same as the Powell's method. Global minimums of doubly excited states have narrow pocket sizes. Hence, rcGA can only approach them and merely reach them. Novel methods for crossover and offspring are necessary for the improvement of the accuracy of rcGA as an optimizer of the VQE method. Although, novel VQE method using the combining method of rcGA and local search method denotes the high accuracy superior to that of BFGS method we demonstrated\cite{2021arXiv210902009W}.
It means that rcGA makes many individuals as parameter sets. They can be used as initial parameter sets of other methods. For example, CVaR-VQE can calculate the optimal folding patterns of proteins by using GA in the process of optimization more accurately than other methods\cite{robert_resource-efficient_2021}. Modifying the initial distribution of individuals also improves the accuracy of final individuals.
We are thinking that there are many local minimum on the initial distribution. We had better introduce local search to change initial distribution. In the future, we want to try to introduce local search to Genetic Algorithm then local search.

\bibliographystyle{apsrev4-2}
\bibliography{temp5}

\end{document}